\documentclass[twocolumn,aps,superscriptaddress]{revtex4}
\usepackage{graphicx}
\usepackage{float}
\usepackage{dcolumn}
\usepackage{bm}
\usepackage{color}
\usepackage{amsmath}
\usepackage{amssymb}
\usepackage{amsfonts}
\usepackage{esint}
\usepackage{times}
\usepackage{xcolor}
\usepackage{phaistos}
\usepackage{braket}
\usepackage{comment}

\usepackage{pdfpages}

\usepackage[colorlinks,linkcolor=blue,anchorcolor=blue,citecolor=blue]{hyperref}

\begin{document}

\title{Tunable coupling of widely separated superconducting qubits: A possible application \\ towards a  modular quantum device }
\author{Peng Zhao}
\email{shangniguo@sina.com}
\affiliation{Beijing Academy of Quantum Information Sciences, Beijing 100193, China}
\author{Yingshan Zhang}
\email{zhangys@baqis.ac.cn}
\affiliation{Beijing Academy of Quantum Information Sciences, Beijing 100193, China}
\author{Guangming Xue}
\email{xuegm@baqis.ac.cn}
\affiliation{Beijing Academy of Quantum Information Sciences, Beijing 100193, China}
\author{Yirong Jin}
\affiliation{Beijing Academy of Quantum Information Sciences, Beijing 100193, China}
\author{Haifeng Yu}
\affiliation{Beijing Academy of Quantum Information Sciences, Beijing 100193, China}

\date{\today}

\begin{abstract}
Besides striving to assemble more and more qubits in a single monolithic quantum device, taking a
modular design strategy may mitigate numerous engineering challenges for achieving large-scalable
quantum processors with superconducting qubits. Nevertheless, a major challenge
in the modular quantum device is how to realize high-fidelity entanglement
operations on qubits housed in different modules while preserving the desired isolation between
modules. In this work, we propose a conceptual
design of a modular quantum device, where nearby modules are spatially separated by centimeters.
In principle, each module can contain tens of superconducting qubits, and can be separately fabricated,
characterized, packaged, and replaced. By introducing a bridge module between nearby qubit modules and
taking the coupling scheme utilizing a tunable bus, tunable coupling of qubits that are housed
in nearby qubit modules, could be realized. Given physically
reasonable assumptions, we expect that sub-100-ns two-qubit gates for qubits housed in nearby
modules which are spatially separated by more than two centimeters could be obtained. In this way, the
inter-module gate operations are promising to be implemented with gate performance
comparable with that of intra-module gate operations. Moreover, with help of through-silicon vias
technologies, this long-range coupling scheme may also allow one to implement inter-module couplers in
a multi-chip stacked processor. Thus, the tunable longer-range coupling
scheme and the proposed modular architecture may provide a promising foundation for solving challenges
toward large-scale quantum information processing with superconducting qubits.
\end{abstract}

\maketitle

\section{Introduction}

After sustained and intense effort in the improvement of qubit performance and
functionality, quantum devices with tens of superconducting qubits have been
realized. This has led to impressive achievements in superconducting quantum-information
processing \cite{Arute2019,Zhu2022,Jurcevic2021}. Nevertheless, the current small-scale
noisy quantum processor is still insufficient to support the pursuit of quantum
advantage (e.g., solving complex problems that are intractable for classical computing) for
practical applications \cite{Preskill2018} and the long-term goal of fault-tolerant quantum
computing \cite{Fowler2012,Martinis2015}. Thus, in addition to striving for further
improvement of qubit performance, focus also begins to shift to the scaling of these
small-scale quantum devices into large-scale quantum systems. Integrating an increasing number
of qubits without scarifying qubit performance, especially in monolithic quantum devices, requires
overcoming several scientific and technical challenges, such as the wiring
problem \cite{Frankea2019,Reilly2019,Martinis2020}, crosstalk \cite{Wenner2011,Barends2014,Huang2021}, and
fabrication yield \cite{Gambetta2017,Brink2018}. To overcome these limitations, various schemes have
been proposed and demonstrated, such as the compact integration of quantum devices with
the classical cryogenic control systems \cite{Leonard2019,McDermott2018,Das2018,Bardin2019,Potocnik2021}, the
three-dimensional (3D) integration technologies \cite{Bejanin2016,Liu2017,Rosenberg2017, Bronn2018,Foxen2018,Spring2020}, and the post-processing of the fabricated qubit devices \cite{Granata2008,Hertzberg2021,Mergenthaler2021}.

From a system integration perspective, building large quantum systems out of smaller modules may
mitigate various challenges faced by the monolithic integration strategy \cite{Brecht2016,Axline2016,Das2018,Zhou2021,Gold2021}.
As in modular devices incorporating several modules, each module can be separately
fabricated, characterized, and replaced. Thus, the fabrication yield of the large modular
system can be improved, and the electromagnetic crosstalk or impact of some correlated errors, such as
caused by the high energy background radiation \cite{Vepsalainen2020,Wilen2021,Cardani2021}, may be
restricted to the module scale. In addition, for modular devices with
larger spatial separation between modules, the vacant space between modules could be
employed to increase the control footprint area for qubit control, and could
even be used to integrate on-chip control electronics \cite{Leonard2019,Das2018,Potocnik2021,Boter2021}.
This allows for a more-compact integration of the qubit device and its classical control
system, thus potentially mitigating the wiring problem \cite{Frankea2019,Reilly2019}. Despite
these appealing features, there is a major challenge in modular quantum devices, i.e., how to
realize fast-speed, high-fidelity entanglement operations across qubits housed in different
modules while keeping adequate physical isolation between modules.

For multiqubit quantum processors, to ensure high-fidelity gate
operations, qubits are generally coupled via a coupler circuit \cite{Chen2014,McKay2016,Yan2018,Mundada2019},
which is employed to mitigate various quantum crosstalks due to parasitic couplings,
such as ZZ crosstalk \cite{Mundada2019}. In this way, the entanglement gate
operations are generally implemented with short-ranged couplings, limiting the spatial
distance between coupled qubits to a few millimeters. In principle, by using these short-range
coupling schemes, inter-module gate operations can be implemented with performance comparable
with that of the intra-module gate operation \cite{Gold2021}. However, in an ideal modular
device, this short-ranged coupling between inter-module qubits is not compatible
with the pursuit of the desired physical isolation between modules.

In this work, we propose a conceptual design of a modular quantum device, where inter-module qubits are
coupled via a bridge module. The inter-module coupling scheme is a natural extension of a previously
proposed scheme utilizing a tunable bus \cite{Zhao2021}, which can be employed for suppressing
parasitic interactions and implementing sub-100-ns Controlled-Z (CZ) gates for inter-module qubits. Given
physical assumptions, we expect that tunable coupling (entanglement
gate operations) of qubits housed in nearby qubit modules, and which are spatially separated by more
than two centimeters could be obtained. Thus, here, it is promising to implement high-fidelity inter-module
gate operations while maintaining the desired physical isolation between modules.

\begin{figure}[tbp]
\begin{center}
\includegraphics[width=8cm,height=10cm]{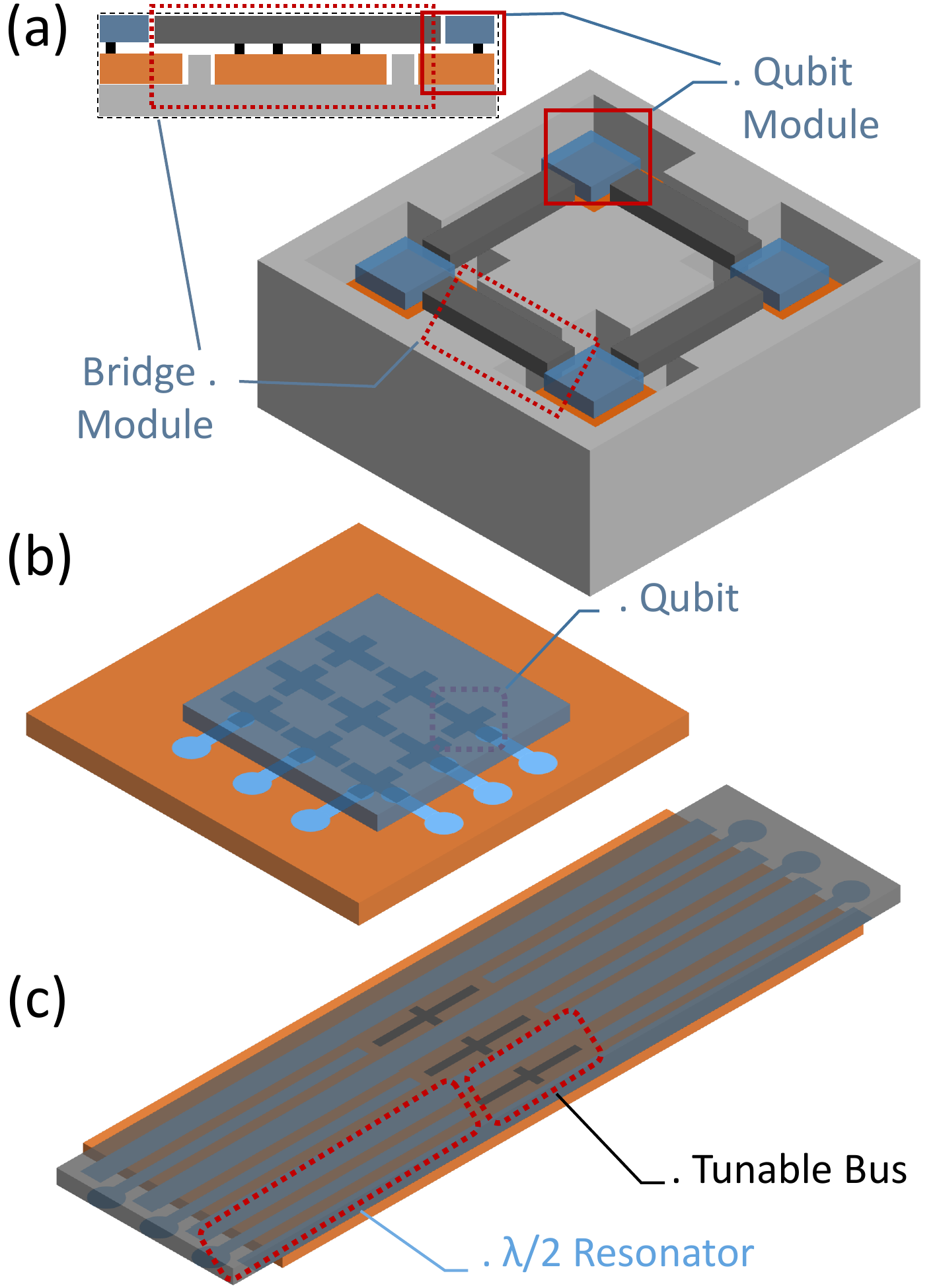}
\end{center}
\caption{Conceptual schematic of a possible modular quantum device utilizing bump-bonding
technologies (not to scale). (a) The modular device comprises two types of functional
modules: qubit modules and bridge modules. Before assembly, each module
can be separately fabricated, packaged, and characterized. After chip assembly and
test, damaged modules due to junction aging of qubits can also be replaced. The inset (outlined with the black
dashed box) shows the cross-sectional view of the modular device. (b) As in traditional monolithic devices utilizing bump-bonding
technologies \cite{Rosenberg2017,Foxen2018}, each qubit module can contain tens of qubits (hosted on the top chip, grey)
and their ancillary circuits (patterned on the carrier chip, orange) for qubit operations. Here, for example,
each qubit module contains a $3\times3$ qubit lattice. (c) The bridge modules
are introduced to mediate coupling between qubits residing in nearby qubit modules, thus
allowing for intra-module entanglement gate operations. The plot presented here shows that each bridge module
comprises three coupler circuits, which are used to mediate inter-qubit coupling between qubits external to each
qubit module. The coupler circuit consists of two $\lambda/2$ CPW resonators which are interconnected
by a tunable bus, and all three elements are hosted on the top chip (grey).}
\label{fig1}
\end{figure}
\section{overview of a modular superconducting quantum processor}

As sketched in Fig.~\ref{fig1}(a), we consider a modular quantum device
comprising two types of functional module: qubit modules and bridge
modules. In principle, before the device assembly, each module can
be separately fabricated, packaged, and characterized. Therefore, only the modules which
function properly during the test, are assembled into the modular
device. This could improve the fabrication yield of the large modular
system.

In addition, after device assembly, damaged modules can arise, such
as due to junction aging of qubits \cite{Koppinen2007,Gates1984,Pop2012,Bilmes2021}. The aging
process is generally attributed to aluminum hydrates present in the junction barrier, which can result from
fabrication residuals \cite{Gates1984,Pop2012}. Moreover, the non-uniformity present in
qubit fabrication \cite{Murray2021} could make the aging process differently over qubits and qubit
modules. To improve the fabrication efficiency of modular devices, it is highly desirable
to replace the damaged module while keeping minimal impacts on the performance of other
functional modules. For this purpose, rather than taking bump-bonding or wire-bonding
technologies for integrating the two-type modules, where qubit modules should have, e.g., 
galvanic connections, with bridge modules, one may prefer underfill
technologies \cite{Das2018}. In this situation, all modules are underfilled to
a chip mount or an additional large carrier chip, as shown in the inset
of Fig.~\ref{fig1}(a), and the coupler circuits housed on the bridge
module are coupled capacitively to the qubits that reside in
the qubit module. This makes it possible to have a modular device with replaceable
modules. However, the underfill epoxies can induce dielectric loss, which
should be minimized to avoid degradation of qubit coherence \cite{Das2018}.

Similar to traditional monolithic quantum devices, each qubit module
can contain tens of qubits and their ancillary circuits
for qubit control, e.g., readout resonators for qubit measurement, and control
lines for gate operations. Fig.~\ref{fig1}(b) shows a typical qubit module utilizing
bump-bonding technologies, where both qubits and their ancillary
circuits can be fabricated by lithographically patterned superconducting metallization
layer, e.g., aluminum, on high-quality substrates, e.g., high resistivity silicon substrates.

As the surface code scheme requires a two-dimensional (2D) lattice of qubits with only nearest-neighbor
couplings \cite{Fowler2012}, here, intra-module qubits can be coupled via the traditional
short-range couplers, and inter-module qubits, i.e., qubits in the outer perimeter of each qubit
module, can be coupled through the coupler circuits housed in bridge modules. Thus, the proposed
modular architecture can be scaled up without any sacrifice of qubit connectivity.
Moreover, the bridge module can decrease qubit density and create vacant space, thus increasing
the control footprint area for qubit control and potentially mitigating the wire problem \cite{Boter2021}.

To couple inter-module qubits, we consider a bus-mediated longer-range coupling scheme. The scheme is an extension
of the recently proposed coupling scheme \cite{Zhao2021}, where two qubits are coupled via a
bus, and tunable $ZZ$ coupling can be achieved through tuning the bus frequency. In addition,
weak qubit-bus coupling, typically with a strength of $20\,\rm MHz$, is adequate for
implementing sub-100-ns CZ gates. Since strong interactions have
been demonstrated for distant qubits coupled to a common half-wave superconducting coplanar
waveguide resonator ($\lambda/2$ CPW resonator) \cite{Sillanpaa2007,Majer2007},
resonator-mediated qubit-bus couplings with strengths of $20\,\rm MHz$ should
be achieved. Thus, we expect that the coupler circuit, which consists of two $\lambda/2$ CPW resonators
connected via a frequency-tunable transmon qubit (acted as a tunable bus), as shown
in Fig.~\ref{fig1}(c), should be a feasible longer-range coupler for realizing
tunable $ZZ$ coupling and implementing sub-100-ns CZ gates for inter-module qubits.

\begin{figure}[tbp]
\begin{center}
\includegraphics[keepaspectratio=true,width=\columnwidth]{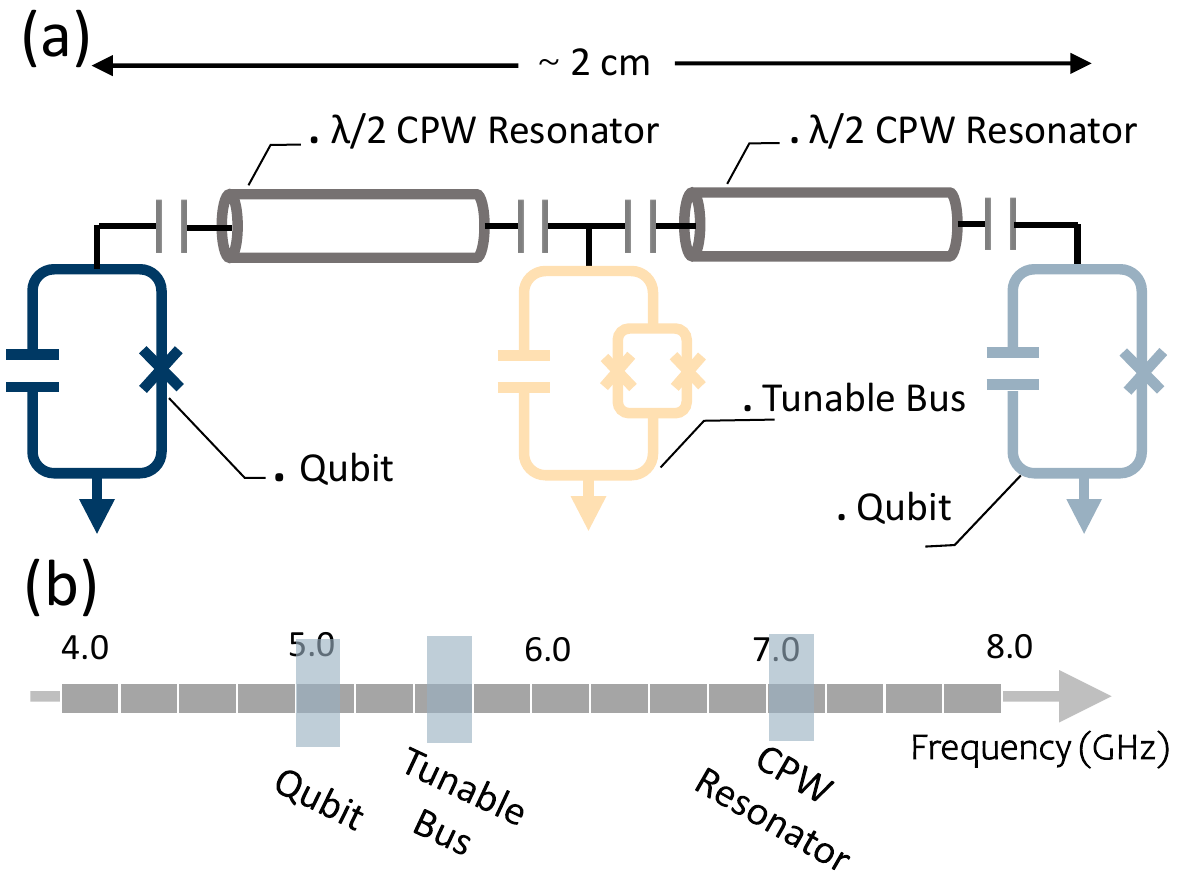}
\end{center}
\caption{Circuit schematic of the proposed coupler circuit for coupling qubits
in nearby qubit modules (not to scale). (a) Two fixed-frequency transmon qubits, that are
spatially separated by $2\,\rm cm$, are coupled together via a coupler circuit. The circuit
consists of two $\lambda/2$ resonators, which are interconnected via
a frequency-tunable transmon qubit (acted as the tunable bus). (b) Frequency arrangement
of the qubit and the coupler circuit (two resonators and one tunable bus).
The qubit frequency is around $5.0\,\rm GHz$, the bus idle frequency is typically $500\,\rm MHz$
above the qubit's, and the fundamental mode frequency of the resonator is about $7.0\,\rm GHz$.}
\label{fig2}
\end{figure}

\begin{figure}[tbp]
\begin{center}
\includegraphics[keepaspectratio=true,width=\columnwidth]{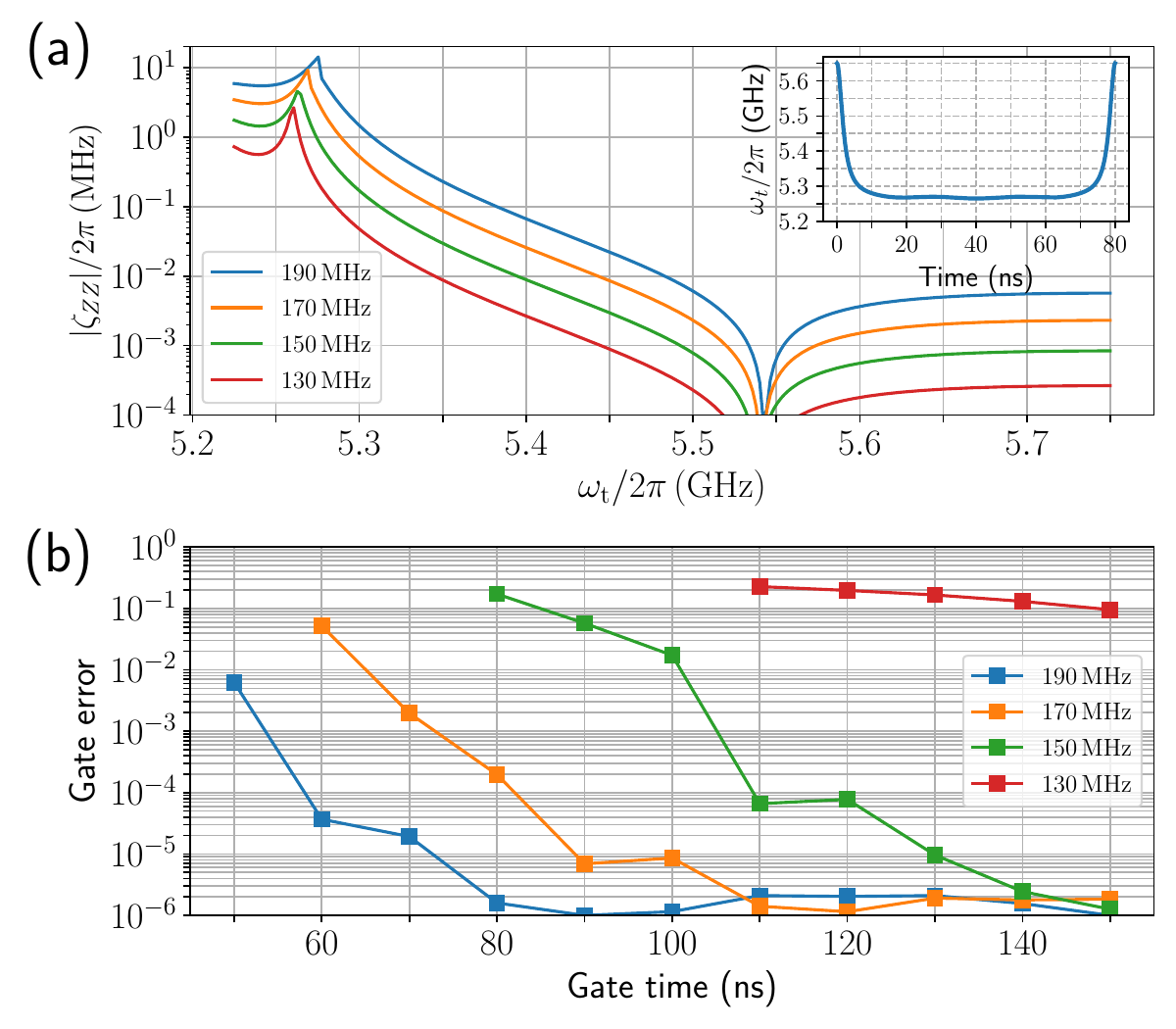}
\end{center}
\caption{Tunable $ZZ$ coupling and its application for implementing two-qubit CZ
gates. The parameters used here are: qubit frequency $\omega_{1(2)}/2\pi=5.0(5.2)\,\rm GHz$,
qubit (bus) anharmonicity $\eta_{1}=\eta_{2}=\eta_{\rm{t}}=\eta$ with $\eta/2\pi=-300\,\rm MHz$,
resonator frequency $\omega_{r1(r2)}/2\pi=7.0(7.2)\,\rm GHz$, and resonator coupling
strength $g_{tk}=g_{rk}=g$ (calculated at $\omega_{i}/2\pi=\omega_{rj}/2\pi
=6.0\,\rm {GHz})$ with $g/2\pi=\{130,\,150,\,170,\,190\}\,\rm MHz$.
(a) $ZZ$ coupling strength $\zeta_{\rm ZZ}$ as a function of the bus
frequency $\omega_{\rm t}$ for the coupled qubit system illustrated in Fig.~\ref{fig2}(a).
The inset shows the typical control pulse shape of the bus
frequency for an 80-ns CZ gate, where the bus frequency is
tuned from the idle point at $5.65\,\rm GHz$ to the working point
near $5.27\,\rm GHz$, and then coming back. (b) CZ gate errors
versus gate time without the consideration of the system decoherence process.}
\label{fig3}
\end{figure}

\section{entanglement operations on qubits housed in nearby modules}
Figure~\ref{fig2}(a) shows the circuit schematic of a coupled two-qubit system,
where two fixed-frequency transmon qubits \cite{Koch2007} (housed in the nearby qubit modules) are
coupled via the proposed coupler circuit (housed in the bridge module). The full
system can be modeled by a chain of five modes including three anharmonic modes (two qubits
and one tunable bus) and two harmonic modes (two resonators) with
nearest-neighboring coupling, described by (hereafter $\hbar =1$)
\begin{eqnarray}
\begin{aligned}\label{eq1}
H=&\sum_{i=1,2,t}\left(\omega_{i}a_{i}^{\dagger}a_{i}+\frac{\eta_{i}}{2}a_{i}^{\dagger}a_{i}^{\dagger}a_{i}a_{i}\right)
+\sum_{j=1,2}\omega_{rj}b_{j}^{\dagger}b_{j}
\\&+\sum_{k=1,2}g_{tk}(a_{t}^{\dagger}+a_{t})(b_{k}+b_{k}^{\dagger})
\\&+\sum_{k=1,2}g_{rk}(a_{k}^{\dagger}+a_{k})(b_{k}+b_{k}^{\dagger}),
\end{aligned}
\end{eqnarray}
where the subscript $i$ labels the anharmonic mode $Q_{i}$ with anharmonicity
$\eta_{i}$ and bare mode frequency $\omega_{i}$, and the subscript $j$ labels
the harmonic mode $R_{j}$ with bare mode frequency $\omega_{rj}$. $a_{i}\,(a_{i}^{\dagger})$ is
the annihilation (creation) operator for the modes $Q_{i}$, and
$b_{j}\,(b_{j}^{\dagger})$ is for modes $R_{j}$. $g_{tk}$ ($g_{rk}$) denotes
the coupling strength between the mode $Q_{t}$ ($Q_{k}$) and the resonator $R_{k}$.

As shown in Fig.~\ref{fig2}(b), here we consider that the qubit frequency is around $5.0\,\rm GHz$,
the bus idle frequency is typically $500\,\rm MHz$ above the qubit's, and the resonator
frequency is about $7.0\,\rm GHz$. Assuming that the dielectric constant of the substrate (silicon)
is $\epsilon_{r} = 11.45$, the $\lambda/2$ CPW resonator with the length $9\,\rm mm$ has the
fundamental mode at $7\,\rm GHz$ \cite{Goppl2008,Pozar2011}. As the typical size of the transom qubit
is about $1\,\rm mm$ \cite{Martinis2020}, we expect that coupling of qubits that are spatially separated
by more than $2\,\rm cm$ can be realized. Thus, given this
long-range coupling scheme, modular devices with large intra-module separation distance could
be achieved, potentially enabling the desired physical isolation between qubit modules.

As mentioned earlier, here, the $ZZ$ coupling of qubits can be controlled
by adjusting the bus frequency. Fig.~\ref{fig3}(a) show the $ZZ$ coupling
strength versus the bus frequency with different resonator coupling
strength. Here, the $ZZ$ coupling strength is defined as (hereafter,
notation $|Q_{1}\,R_{1}\,Q_{t}\,R_{2}\,Q_{2}\rangle$ denotes the full system
state, and when confined to qubit subspace, notation $|Q_{1}\,Q_{2}\rangle$ is used)
\begin{eqnarray}
\begin{aligned}\label{eq2}
\zeta_{\rm ZZ}\equiv(E_{11}-E_{10})-(E_{01}-E_{00}),
\end{aligned}
\end{eqnarray}
and can be obtained numerically by dialogizing the full system Hamiltonian in Eq.~(\ref{eq1}).
In Eq.~(\ref{eq2}), $E_{jk}$ denotes the eigenenergy of the full system associated with
eigenstate $|\tilde{jk}\rangle$, which is adiabatically connected to the bare
state $|j000k\rangle$ \cite{Ghosh2013}.

From the results shown in Fig.~\ref{fig3}(a), one can find that the $ZZ$ coupling
can be suppressed below $10\,\rm kHz$ by tuning the bus frequency above $5.5\,\rm GHz$, thus
high-performance single-qubit control can be realized with $ZZ$-suppression. Furthermore,
when the resonator coupling strength $g$ takes values larger than $170\,\rm MHz$, $ZZ$ coupling
with strength above $10\,\rm MHz$ can be achieved with the bus frequency near $5.27\,\rm GHz$,
thus sub-100-ns CZ gates should be obtainable. Here, we consider that the CZ gate is
implemented by tuning the bus frequency according to a fast adiabatic
control pulse \cite{Martinis2014}. The typical control pulse shape of the
bus frequency can be found in the inset of Fig.~\ref{fig3}(a), where the bus frequency is
tuned from the idle point (at $5.65\,\rm GHz$, where ZZ coupling is suppressed
below $10\,\rm kHz$) to the working point (near $5.27\,\rm GHz$, where the ZZ coupling takes
its maximum value, i.e., about $10\,\rm MHz$), and then coming back. The detailed procedure
for tuning up the fast-adiabatic CZ gate and characterizing the intrinsic CZ gate
performance (i.e., in the absent of decoherence) can be found
in Ref.\cite{Zhao2021}.

To evaluate gate performance, Figure~\ref{fig3}(b) shows
the CZ gate error as a function of the gate time without
the consideration of the system decoherence process \cite{Pedersen2007}.
One can find that with the resonator coupling strength above $170\,\rm MHz$,
high-fidelity sub-100-ns CZ gates can be achieved. This suggests that with
this tunable coupling scheme, the inter-module gate operations are promising
to be implemented with gate performance comparable with that of
intra-module gate operations.

\begin{figure}[tbp]
\begin{center}
\includegraphics[keepaspectratio=true,width=\columnwidth]{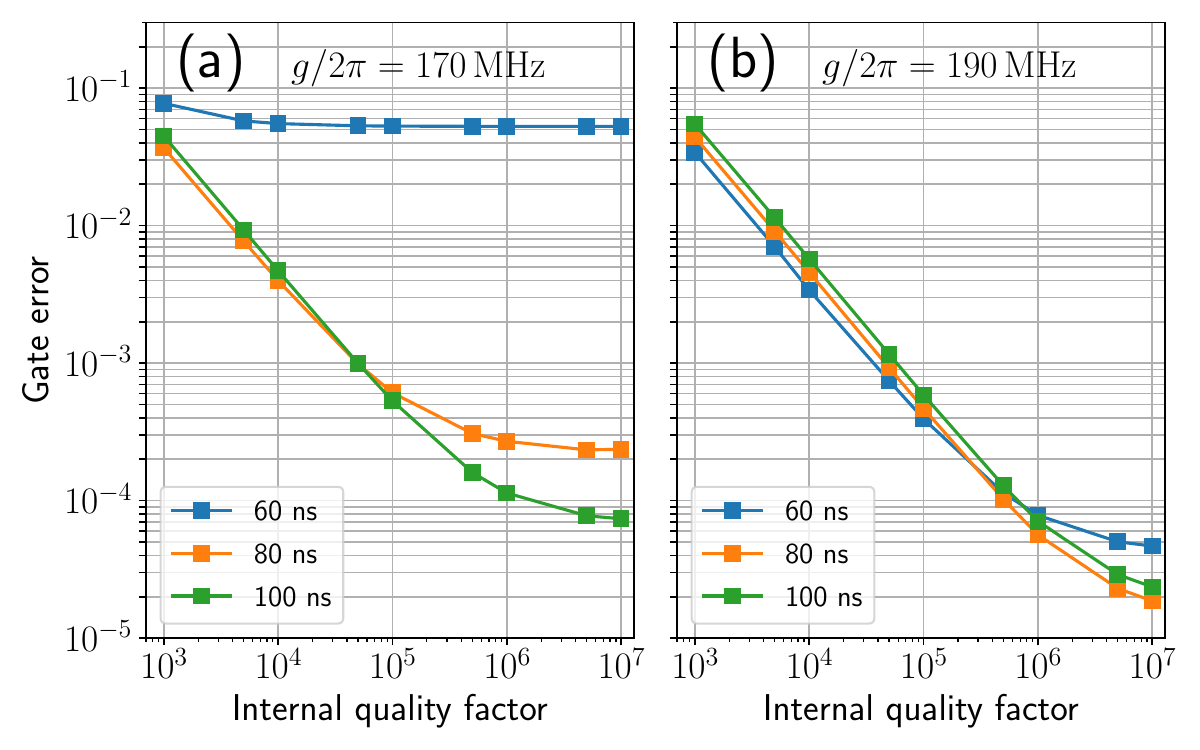}
\end{center}
\caption{CZ gate error versus the internal quality factor of the resonator.
The resonator coupling strength is $g/2\pi=170\,\rm MHz$ for (a),
and $g/2\pi=190\,\rm MHz$ for (b), and the decay rate of the resonator
is $\kappa_{j}=\omega_{rj}/Q_{j}$ (here, the internal quality factors of the two resonators take the
same value). The other system parameters are the
same as in Fig.~\ref{fig3}. }
\label{fig4}
\end{figure}

In practical implementations, compared with the original scheme, one
may expect that besides the decoherence processes of the qubits and the bus, the
difficulties to face with the present scheme depend on one additional limitation, i.e., the
resonator decay process. Here, to study the influence of the resonator decay process
on the CZ gate performance, the Lindblad master equation is employed. To be more
specific, by considering the resonator decay process, the master equation can be expressed as
\begin{eqnarray}
\begin{aligned}\label{eq3}
\frac{d\rho}{dt}=-i[H,\rho]+\sum_{j=1,2}\kappa_{j}\mathcal{L}[b_{j}],
\end{aligned}
\end{eqnarray}
where the Hamiltonian $H$ is given in Eq.~(\ref{eq1}), $\rho$ is the density
matrix of the system, $\mathcal{L}[b_{j}]=b_{j}\rho b_{j}^{\dagger }-b_{j}^{\dagger }b_{j}\rho/2-\rho b_{j}^{\dagger }b_{j}/2$
describes the resonator decay terms, $\kappa_{j}$
denotes the photon decay rate of resonator $R_{j}$. The average gate fidelity of the
CZ gate under the resonator decay process is defined as \cite{Wood2018}
\begin{eqnarray}
\begin{aligned}\label{eq4}
F=\frac{d\,F_{p}+1-L_{1}}{d+1},
\end{aligned}
\end{eqnarray}
where $d$ denotes the dimension of the computational subspace of the
system, $L_{1}$ represents the leakage of the gate operation, and $F_{p}$ is
the process fidelity of the implemented CZ gate. The process fidelity $F_{p}$ can be
obtained by numerically performing quantum process tomography of the implemented CZ
gate based on the master equation given in Eq.~(\ref{eq3}). Similarly, the leakage $L_{1}$
can be obtained by solving the master equation with the system initialized in
different computational states \cite{Wood2018}.

Figure~\ref{fig4} shows the CZ gate error $1-F$ versus the internal quality
factor of the resonator. one can find that sub-100-ns CZ
gates with gate errors below $0.01$ ($0.001$) can be achieved with the
resonator internal quality factor exceeding $5\times10^{3}$ ($5\times10^{4}$). Given
the state-of-the-art results, planar superconducting CPW resonators with internal
quality factors above $10^{6}$ have been demonstrated \cite{Megrant2012,Calusine2018,Gao2022}.
Therefore, the result shown in Fig.~\ref{fig4} suggests that with current technologies, the
gate error from the resonator decay process is promising to be pushed below $0.001$.

In addition, in the presence of decoherence, there generally exists two gate error
channels, i.e., decoherence error due to the resonator decay and coherence error
including leakage resulting from non-adiabatic transitions. The two channels
favor opposite gate strategies, i.e., short gate times for mitigating decoherence
error and longer gate times for suppressing coherence error. The tradeoff
between the two types of error can explain that the gates with longer gate times
do not always show better performance, as shown in Fig.~\ref{fig4}. Moreover, by comparing
the results in Fig.~\ref{fig4}(a) and~\ref{fig4}(b), one can find that although larger resonator
coupling can decrease the coherent error, as also shown in Fig.~\ref{fig3}(b), it can, on the
contrary, increase the decoherence error through the resonator decay process. This is to be
expected, as larger resonator coupling can result in a strong state hybrid between the qubit
and the resonator, thus qubit relaxation through the resonator can become
more serious \cite{McKay2016}.

\section{discussion and conclusion}

In summary, the work aims to show the possibility of solving challenges toward
large-scale quantum devices with tunable longer-range coupling. We have shown
that with the proposed tunable longer-range coupler, a modular quantum device with
fast-speed, high-fidelity inter-module gate operations and desired physical
isolation could be obtained. This could give rise to a promising foundation for
mitigating several scientific and technical
challenges toward large-scale superconducting quantum computing. Though, in practice, the feasibility of the
proposed modular design may be limited because of several relevant engineering
challenges, e.g., the assembly accuracy of modules \cite{Gold2021}, and the parasitic
electromagnetic modes within the device package \cite{Wenner2011,Huang2021}, another
goal of this work is to encourage further experimental and theoretical research
in incorporating long-range inter-qubit coupling into scalable
quantum information processing with superconducting qubits \cite{Tremblay2021}.

While in the present work, the introduced longer-range coupler is employed for realizing longer-range
tunable $ZZ$ coupling and CZ gates, in the supplementary material,
we further show that the coupler could also be used for implementing other types of two-qubit gate.
We will show that by setting the bus frequency at $5.5\,\rm GHz$, the $ZZ$ coupling has
been suppressed heavily, as shown in Fig.~\ref{fig3}(a), meanwhile, the transversal ($XY$) coupling between qubits can still be
maintained \cite{Zhao2021b}. Thus, for qubits coupled via the proposed coupler, all-microwave-controlled
cross-resonance gates (or CX gates) can be realized with $ZZ$ suppression \cite{Kandala2021}.

\begin{acknowledgments}

We would like to thank Xuegang Li and Junling Long for their helpful
discussions and comments. Numerical analysis of the resonator decay process
on gate performance were performed using the Quantum Toolbox in
Python (QuTiP) \cite{Johansson2012}. This work was partly supported by the National
Natural Science Foundation of China (Grants No.11890704, No.12004042), the Beijing
Natural Science Foundation (Grant No.Z190012), and the Key-Area Research and
Development Program of Guang Dong Province (Grant No. 2018B030326001).

\end{acknowledgments}

\appendix

\clearpage



\section*{Supplementary material}

# Supplementary material for "Tunable coupling of widely separated superconducting qubits: A possible application towards a modular quantum device"

Peng Zhao,[1, *] Yingshan Zhang,[1, †] Guangming Xue,[1, ‡] Yirong Jin,[1] and Haifeng Yu[1]

[1]*Beijing Academy of Quantum Information Sciences, Beijing 100193, China*
(Dated: May 26, 2022)

### Cross-resonance gate

In the main text, we show that the introduced longer-range coupler can be employed for realizing longer-range tunable $ZZ$ coupling and CZ gates by controlling the bus frequency. Here, we further show that the proposed coupler can also be used to implement cross-resonance gates [1]. Cross-resonance (CR) gate is an all-microwave-controlled two-qubit gate, which is realized by driving the control qubit at the frequency of the target qubit, and has been demonstrated as a leading two-qubit gate scheme for qubit architectures with fixed-frequency qubits [2]. One of the major parasitic interactions, that degrade the performance of CR gates, is the parasitic $ZZ$ coupling [1, 3]. Since the implementation of CR gates is based on the transversal ($XY$) coupling between qubits, it is highly desirable to suppress the parasitic ZZ coupling while keeping adequate $XY$ coupling strength for implementing CR gates [4, 5]. From the point of view of perturbation theory, the $XY$ and $ZZ$ coupling are generally enabled by different virtual transition paths, thus the suppression of $ZZ$ coupling does not always mean that the $XY$ coupling is also suppressed [4]. This suggests that one can maintain the XY interaction while suppressing ZZ coupling, thus XY-interaction-based two-qubit gates, such as iSWAP gates or CR gates, can be realized with ZZ suppression [4, 5].

### CX gate

In the following discussion, we will show that CR gates can be realized with ZZ suppression for two qubits coupled via the proposed longer-range coupler (see Fig.2 of the main text). For this system, by setting the bus frequency above $5.5\,\mathrm{GHz}$, the $ZZ$ coupling can be suppressed below $10\,\mathrm{kHz}$ (see Fig.3(a) of the main text). Thus, we consider that in the coupled qubit system, the bus takes the fixed frequency at $5.5\,\mathrm{GHz}$, suppressing the parasitic $ZZ$ coupling for implementing CR gates. The resonator coupling strength $g$ is $190\,\mathrm{MHz}$, and other system parameters are the same as in Fig.3(a) of the main text.

Here, we intend to realize CX (CNOT) gate using the CR gate scheme, i.e., driving the control qubit $Q_2$ at the frequency of the target qubit $Q_1$. In addition, besides the CR drive applied to $Q_2$, here we consider that an additional cancellation drive is

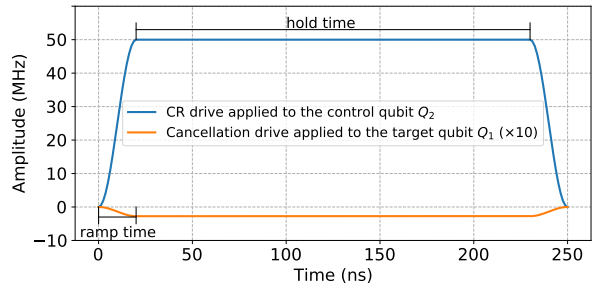

FIG. S1: Typical control pulse with cosine-shaped ramp for implementing CX (CNOT) gates, where the full width at maximum is defined as the hold time.

---

*Electronic address: shangniguo@sina.com
†Electronic address: zhangys@baqis.ac.cn
‡Electronic address: xuegm@baqis.ac.cn

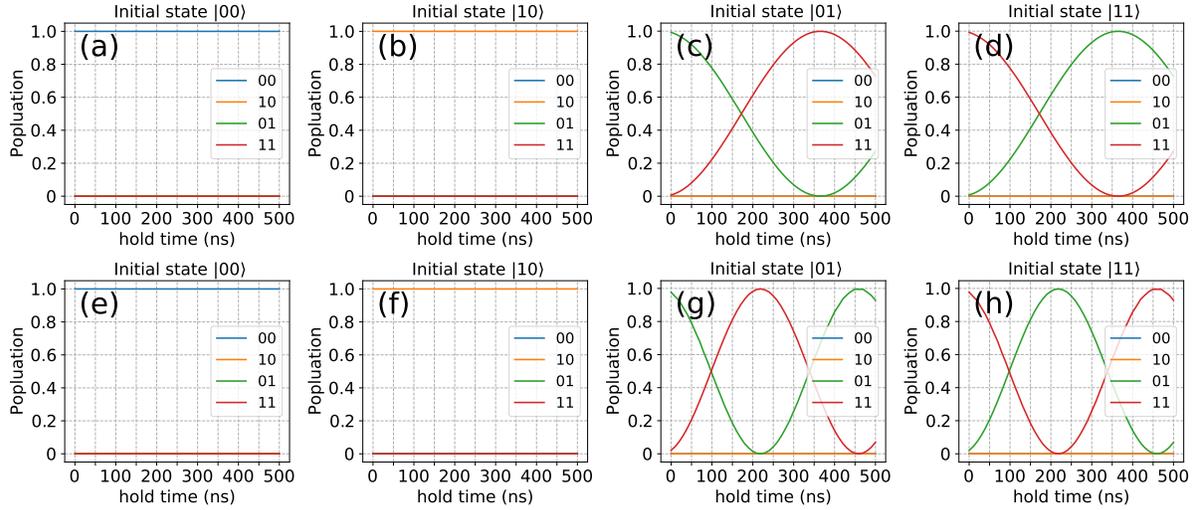

FIG. S2: Populations versus the hold time for four initial states with the CR drive and the cancellation drive applied to $Q_2$ and $Q_1$, respectively. The typical pulse shape is given in Fig. S1. Here, the ramp time is 20 ns. (a-d) the amplitude of the CR drive is $\Omega_2/2\pi = 50\,\text{MHz}$ and the amplitude of the cancellation drive is $\Omega_1 = -0.0054\Omega_2$. (e-h) the amplitude of the CR drive is $\Omega_2/2\pi = 100\,\text{MHz}$ and the amplitude of the cancellation drive is $\Omega_1 = -0.0058\Omega_2$.

applied to $Q_1$ [2]. The system Hamiltonian under microwave drives can be expressed as

$$H_{full} = H + H_d,\ H_d = \sum_{i=1,2} \Omega_i(a_i^\dagger + a_i)\cos(\omega_d t),$$

where $H$ is the undriven system Hamiltonian given in Eq.(1) of the main text, $\Omega_{1(2)}$ denotes the amplitude of the cancellation drive (CR drive) applied to $Q_{1(2)}$, and $\omega_d$ is the drive frequency, i.e., the (dressed) frequency of $Q_1$.

For the CX gate calibration, the amplitude of the cancellation drive is chosen to ensure no operation on the target qubit $Q_1$ when the control qubit $Q_2$ is in state $|0\rangle$ [6]. Thus, up to single-qubit phase compensation (single-qubit phase gate), the CX gate can be implemented by carefully choosing the amplitude of the CR drive and the gate time. The typical pulse shape for the CX gate is shown in Fig. S1, where the pulse has cosine-shaped ramps, and the full width at maximum is defined as the hold time.

Figure S2 shows the qubit population versus the hold time during the gate operations. One can find that with the cancellation drive, there is indeed no operation on $Q_1$ when $Q_2$ is in state $|0\rangle$. On the contrary, when $Q_2$ is in state $|1\rangle$, the population of $Q_1$ shows a normal Rabi oscillation. Thus, by choosing optimal hold times, CX gates can be obtained.

As shown in Fig. S2(a-d), for a CR drive with $\Omega_2/2\pi = 50\,\text{MHz}$, a CX gate can be obtained with hold time of 365 ns, and the gate fidelity is 0.9996 [7]. The gate time could be shorter by increasing the CR drive amplitude. For example, as shown in Fig. S2(e-h), where the amplitude is 100 MHz, a CX gate can be achieved with the hold time of 220 ns and the gate fidelity is 0.9988. The increased gate error can be reduced by increasing the ramp time [6]. When increasing the ramp time to 25 ns, a higher-fidelity (0.9996) CX gate can be obtained with the hold time of 213 ns.

---



\end{document}